\documentclass[12pt]{article}

\usepackage{amsthm,graphicx,cite,lscape,float,multirow}
\usepackage{epsf,latexsym}
\usepackage{amssymb,amsmath}

\newcommand{\nhatpa}{\hat{\rm n}_{\mbox{\scriptsize $\parallel$}}}
\newcommand{\nhatp}{\hat{\rm n}_{\perp}}
\newcommand{\fp}{\vec{f}_{\mbox{\scriptsize $\parallel$}}}
\newcommand{\fpm}{f_{\mbox{\scriptsize $\parallel$}}}

\begin{document}

\title{\bf New insights into the capillary retention force and 
the work of adhesion}

\author{Rafael de la 
Madrid,\footnote{E-mail: \texttt{rafael.delamadrid@lamar.edu}} \ 
Huy Luong,\footnote{Current address: Sage Automation, Beaumont, TX 77705}
\ Jacob Zumwalt \\
\small{\it Department of Physics, Lamar University,
Beaumont, TX 77710} }

\date{\small{December 22, 2021}}



\maketitle

\begin{abstract}
  \noindent 
We calculate the normal capillary retention force that anchors a drop to 
a solid surface in the direction perpendicular to the surface, and study
the relationship between such force and the Young-Dupr\'e work of 
adhesion. We also calculate the work necessary to 
create or destroy a patch of solid-liquid interface 
by moving the triple line on a solid substrate. We argue that
when the capillary number is small and a drop is sliding on a surface at 
constant speed, the lateral retention force is the major source of energy 
dissipation, whereas viscous dissipation plays a minor role. 
\end{abstract}


\section{Introduction}
\label{sec:introduction}

The adhesion of droplets to solid surfaces has been studied extensively
due to its intrinsic scientific interest and its wide range of important
applications, such as self-cleaning, microfluidics, ink-jet printing,
mist eliminators, and carbon dioxide capture, to name just a few. Three
very important quantities that characterize the adhesion properties
of a solid-liquid pair are the lateral retention force
$\fp$ (which quantifies the opposition to lateral motion of a drop moving 
on a solid substrate), the normal force of adhesion $\vec{f}_{\perp}$
(which quantifies the resistance of a drop to be detached from a surface),
and the Young-Dupr\'e work of adhesion ${\sf w}_{\rm adhesion}$
(which quantifies the energy per unit area necessary to detach a drop from 
a surface without changing the shape of the drop).

Although the lateral retention force has been studied extensively (see 
Refs.~\cite{FURMIDGE,DUSSAN,EXTRAND,CARRE,ELSHERBINI,ANTONINI,
TADMOR13,CONINCK,GRIEGOS1,GRIEGOS2,GRIEGOS3,TAFRESHI18d,LANGMUIR-RAFA,DUNLOP} 
and references therein), its normal counterpart $\vec{f}_{\perp}$ has not 
received nearly as much attention~\cite{DUNLOP,DEGENNES,PELLICER,LESTER,SHA1,SHA2,SHA3,SHA4,SHA5,SHA6,SHA7,SHA8,TADMOR17,EXTRAND17,TAFRESHI18a,TAFRESHI18b,TAFRESHI18c,TAFRESHI21}, in spite of the fact that it plays an essential role
in many situations. A familiar situation is a pendant drop, which must be 
anchored to the surface by a force that is normal to the surface. In this
paper, our first
objective is to derive an expression for $\vec{f}_{\perp}$ using the same
kind of arguments used to derive the expression for $\fp$. The resulting
expression for $\vec{f}_{\perp}$ turns out to be essentially the same as that of 
Refs.~\cite{DUNLOP,TAFRESHI18a}.

The expression for the Young-Dupr\'e work of adhesion has been known for 
over 150 years~\cite{DUPRE}, and it is part of the standard 
theory of wetting~\cite{DAVIES,ADAMSON,BERG,SCHRADER}. However, its precise 
measurement has proved to be elusive. In a recent experiment~\cite{TADMOR17}, 
Tadmor {\it et al.}~presented a new way to measure 
${\sf w}_{\rm adhesion}$. That experiment was re-analyzed by
Extrand~\cite{EXTRAND17}. Our second objective is to use the
expression for $\vec{f}_{\perp}$ to provide a new theoretical description of
that experiment. We will show that the normal retention force per unit of 
triple line is different from the Young-Dupr\'e work of adhesion, and
we will compare the (apparent) contact angles predicted by $\vec{f}_{\perp}$ with
those of Refs.~\cite{TADMOR17,EXTRAND17}.

The work of adhesion quantifies the energy needed to detach a drop
from a surface when the drop is in thermodynamic equilibrium. However, the 
energy needed
to move the triple line along the surface in dynamical situations where the drop
may not be in thermodynamic equilibrium has been rarely 
discussed~\cite{FURMIDGE}. Our third objective is to calculate
the work done (i.e., energy dissipated) by the lateral retention force 
when the triple line is moving on a surface. We will introduce two new 
quantities, the advancing (${\sf w}_{\rm a}$) and the receding 
(${\sf w}_{\rm r}$) works of adhesion, which arise from the energy
dissipated by $\fp$ as the triple line moves on a surface. The parameter
${\sf w}_{\rm a}$ (${\sf w}_{\rm r}$) will quantify the 
work needed to create (destroy) a unit of solid-liquid contact area 
by expanding (contracting) the triple 
line. The expressions for ${\sf w}_{\rm a}$ and 
${\sf w}_{\rm r}$ will be derived using purely mechanical arguments (work 
done by the lateral retention force) and can be applied to dynamical
situations where the drop may not be in thermodynamic equilibrium. We will 
see that, even though
${\sf w}_{\rm a}$ and ${\sf w}_{\rm r}$ are not thermodynamic equilibrium
quantities, their expressions resemble that of ${\sf w}_{\rm adhesion}$. 

Our fourth objective is to use the expressions for ${\sf w}_{\rm a}$ and 
${\sf w}_{\rm r}$ to re-derive Furmidge's expression~\cite{FURMIDGE} for the 
sliding work of adhesion, ${\sf w}_{\rm sliding}$, and to argue that, when
a drop is moving at constant speed and the capillary number is small, viscous 
dissipation can be neglected
and $\fp$ is the main source of energy dissipation. 

Our overall objective is that the results of this paper contribute to 
a complete description of the adhesion properties of droplets in terms
of forces and energy, in both the parallel and the normal directions to 
the surface.

The structure of the paper is as follows. In Sec.~\ref{sec:lrfo}, we
recall the main arguments used to derive the expression for the lateral
retention force. We also 
recall the expressions for the works of adhesion (Young-Dupr\'e) and 
cohesion. In Sec.~\ref{sec:nfoa}, we calculate $\vec{f}_{\perp}$. In 
Sec.~\ref{sec:arwa}, we introduce
${\sf w}_{\rm a}$ and ${\sf w}_{\rm r}$. In Sec.~\ref{ref:lwoa}, we re-derive 
the expression for ${\sf w}_{\rm sliding}$. In Sec.~\ref{sec:exsect}, we discuss 
the details of a simple experiment that will be used to test the 
expressions for ${\sf w}_{\rm a}$, ${\sf w}_{\rm r}$ and
${\sf w}_{\rm sliding}$. In Sec.~\ref{sec:nfoad}, 
we analyze the relationship between $\vec{f}_{\perp}$ and the 
Young-Dupr\'e work of adhesion, and provide a new theoretical
description of the experiment of Ref.~\cite{TADMOR17}. In 
Sec.~\ref{sec:lwoad}, we obtain the experimental values
of ${\sf w}_{\rm a}$, ${\sf w}_{\rm r}$, ${\sf w}_{\rm sliding}$ and 
${\sf w}_{\rm adhesion}$, and we 
infer from such values that viscous dissipation can be neglected in our 
solid-liquid system. Finally,
Sec.~\ref{sec:conclusions} contains our conclusions.

\section{Theoretical section}

\subsection{The lateral retention force, and the Young-Dupr\'e work
of adhesion} 
\label{sec:lrfo}

To better understand how the expressions for $\vec{f}_{\perp}$,
${\sf w}_{\rm a}$ and ${\sf w}_{\rm r}$ arise, it is useful to first recall a
derivation~\cite{DUSSAN,LANGMUIR-RAFA} of the lateral retention force $\fp$.

\begin{figure}[h!]
\begin{center}
              \epsfxsize=10cm
              \epsffile{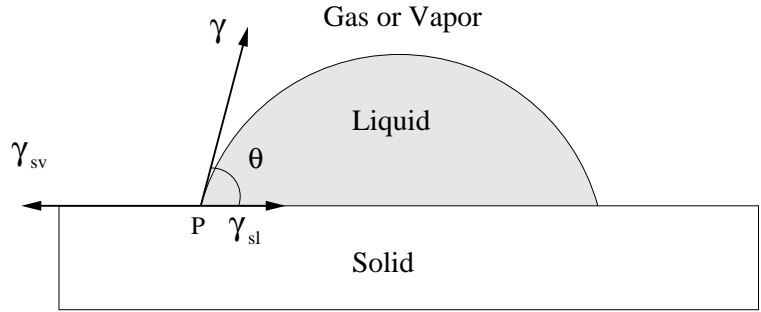}
\end{center}                
\caption{Surface tensions and contact angle $\theta$ at a generic point P of
the triple line. The infinitesimal section $dl$ (not shown in the figure)
is perpendicular to the page at point P.}
\label{fig:stcca}
\end{figure}

Let us consider a drop on top of a flat surface. Let us denote
the liquid-vapor, solid-vapor and solid-liquid surface tensions by
$\gamma =\gamma _{\rm lv}$, $\gamma _{\rm sv}$ and $\gamma _{\rm sl}$, 
respectively. Such surface tensions act on a given infinitesimal section 
$dl$ of the triple line at point P 
as shown in Fig.~\ref{fig:stcca}.

When the contact angle $\theta$ is the Young, equilibrium angle 
$\theta_{\rm Y}$, the net force on the element $dl$ is zero. This balance of 
forces leads to the Young equation,
\begin{equation}
       \gamma \cos \theta_{\rm Y} +\gamma_{\rm sl}-\gamma_{\rm sv}=0 \, .
       \label{youngeq}
\end{equation}
When the contact angle 
$\theta$ at $dl$ is not $\theta _{\rm Y}$, the
forces due to the surface tensions do not cancel each 
other~\cite{DEGENNES},
\begin{equation}
       \gamma \cos \theta + \gamma_{\rm sl}-\gamma_{\rm sv} \neq 0 \, ,
        \label{sost}
\end{equation}
and therefore there is a non-zero capillary force per unit of length
acting at point P in the direction
parallel to the surface. If we denote by $\nhatpa$ the unit vector
that is parallel to the surface, perpendicular to the triple line and 
pointing outwardly, then such
capillary force per unit of length is given by
\begin{equation}
       (-\gamma \cos \theta -\gamma_{\rm sl}+\gamma_{\rm sv})\, \nhatpa \, .
        \label{cfpuol}
\end{equation}
However, if the infinitesimal element $dl$ at point P does not move,
there must be a force that cancels that in Eq.~(\ref{cfpuol}), and therefore
has the same magnitude but opposite direction. The force that cancels
that in Eq.~(\ref{cfpuol}) is the lateral retention 
force per unit of length,
\begin{equation} 
       \frac{\rm retention \ force}{\rm length \ of \ triple \ line} =    
(\gamma \cos \theta +\gamma_{\rm sl}-\gamma_{\rm sv})\, \nhatpa=
           \gamma (\cos \theta - \cos \theta _{\rm Y}) \, \nhatpa \, ,
        \label{magretforvec}
\end{equation}
where in the last step we have used the Young equation. Hence, the 
(infinitesimal) lateral retention force on an (infinitesimal) segment $dl$ 
of triple line is
\begin{equation}
    \gamma (\cos \theta - \cos \theta _{\rm Y}) dl 
                      \, \nhatpa \, .
        \label{larfodlt}
\end{equation}
The total retention force on a drop can be calculated by integrating
Eq.~(\ref{larfodlt}) along the triple line. When the 
triple line has a quasi-rectangular 
shape, and when the advancing (receding) contact angle is constant along the
advancing (receding) edge of the drop, integration of 
Eq.~(\ref{magretforvec}) along the triple line yields~\cite{DUSSAN} exactly
Furmidge's expression~\cite{FURMIDGE},
\begin{equation}
     \fpm= \gamma w (\cos \theta_{\rm r}-\cos \theta _{\rm a}) \, ,
     \label{trf}
\end{equation}
where $w$ is the width of the drop, and $\theta_{\rm a}$ ($\theta_{\rm r}$) is
the angle that the liquid-air interface makes with the solid 
at the advancing (receding) edge of the drop.

In principle, the derivation of Eq.~(\ref{larfodlt}) applies when the drop is 
not moving. When the drop is moving, an additional source of opposition to
lateral motion arises from viscous dissipation~\cite{DEGENNES}. However,
even though the drop is moving, there is still an imbalance of 
forces~\cite{DEGENNES}, and we expect Eq.~(\ref{larfodlt}) to hold, although 
now $\theta$ is a dynamical contact angle~\cite{DEGENNES}. In fact, if 
the capillary force of Eq.~(\ref{larfodlt}) suddenly disappeared when 
the drop starts to move, right after the onset of the motion, when the 
speed is very low, the viscous force would be very small, and the drop would 
first accelerate dramatically and later
slow down when viscous dissipation sets in. However, it is well known
that when a drop starts to move due to a slowly increasing lateral force, the 
drop speeds up slowly, and therefore there must be a retention force at 
low speeds that is not due to viscous dissipation. Hence, when the drop is 
in motion, we will assume
that Eq.~(\ref{larfodlt}) accounts for the capillary contribution to the
lateral retention force. In addition, in situations where the capillary
number is low, we expect
that viscous dissipation can be neglected and that the (capillary) retention
force of Eq.~(\ref{larfodlt}) provides the main opposition to the motion 
of the drop.

The work per unit area necessary to detach a drop from a solid substrate 
without changing the shape of the drop is given 
by~\cite{DAVIES,ADAMSON,BERG,SCHRADER}
\begin{equation}
      {\sf w}_{\rm adhesion}=\frac{W_{\rm adhesion}}{A}=\gamma +\gamma _{\rm sv} -
           \gamma_{\rm sl} \, ,
       \label{defwofa}
\end{equation}
where $W_{\rm adhesion}$ is the total work of adhesion, $A$ is the area of the 
solid-liquid interface, and ${\sf w}_{\rm adhesion}$ is the work of adhesion 
per unit area~\cite{NOTE1}. Thanks to the Young equation,
Eq.~(\ref{defwofa}) can be written as the Young-Dupr\'e equation,
\begin{equation}
      {\sf w}_{\rm adhesion}=\frac{W_{\rm adhesion}}{A}= \gamma (1+\cos \theta_{\rm Y}) \, .
       \label{wofa}
\end{equation}
When we separate a liquid from itself, Eq.~(\ref{defwofa}) must be replaced 
by the work of cohesion~\cite{DAVIES},
\begin{equation}
      {\sf w}_{\rm cohesion}=\frac{W_{\rm cohesion}}{A}= 2 \gamma  \, ,
       \label{woalfl}
\end{equation}
where $A$ is the area of the newly created liquid-vapor interface.


\subsection {The normal retention force} 
\label{sec:nfoa}

In the previous section, we used classic arguments~\cite{DUSSAN,DEGENNES} to 
derive the well-known expression of the lateral retention force. We are now 
going to use the same arguments~\cite{DEGENNES} to calculate
the normal retention force.

Let us assume that there is an external force in the vertical direction 
pulling the drop away from the surface. Even if there was no gravity, the 
drop would not be detached from the surface unless the pulling force is 
strong enough, because 
there is a normal force of adhesion that pins the drop to the surface. From 
Fig.~\ref{fig:stcca}, we can see that $\gamma \sin \theta$ is the only 
component 
of surface tension acting in the normal direction. Because 
the infinitesimal element $dl$ does 
not move in the normal direction, there must be a force pointing in the 
opposite direction and pinning the triple line to the 
surface~\cite{DEGENNES}. Hence,
the normal force of capillary adhesion per unit length should be
\begin{equation}
   \frac{d\vec{f}_{\perp}}{dl}= \gamma \sin \theta\, \nhatp \, ,
   \label{difoa}
\end{equation}
where $\nhatp$ is a unit vector perpendicular to the surface 
and pointing away from the drop. To obtain 
the total retention force that 
pins the drop in the normal direction, we just need to integrate 
Eq.~(\ref{difoa}) along the triple line,
\begin{equation}
   \vec{f}_{\perp} = \oint \gamma\, \sin \theta \, dl\,  \nhatp \, .
   \label{difoat}
\end{equation}
For the particular case that the contact angle remains constant along the
triple line and that the triple line is a circumference of radius $r$, 
Eq.~(\ref{difoat}) yields 
\begin{equation}
   \vec{f}_{\perp} = 2\pi r \gamma\, \sin \theta  \,  \nhatp \, .
   \label{difoasc1}
\end{equation}
Not surprisingly, Eq.~(\ref{difoasc1}) has a resemblance with
Tate's law~\cite{ADAMSON}.

It is important to realize that the magnitude of the actual external force 
$f_{\rm d}$ necessary to detach a drop from a surface is in general different
from $f_{\perp}$. For example, the force necessary to detach a sessile
drop is different than for a pendant drop, because weight opposes
(helps) the detachment of a sessile (pendant) drop. In addition, even
in the absence of gravity, the Laplace 
pressure~\cite{TAFRESHI18a,TAFRESHI18b,TAFRESHI18c,TAFRESHI21,DUNLOP}
produces a force in the normal direction that needs to be taken
into account. However, even though $f_{\perp}$ is different from 
$f_{\rm d}$, in this paper we will refer to $\vec{f}_{\perp}$ as the 
normal force of adhesion for two reasons. First, 
$\vec{f}_{\perp}$ is the only force that is exerted by the solid on the triple
line and that tries to anchor the drop on the solid in the normal 
direction. The force produced by the pressure at the solid-liquid contact area
points away from the surface~\cite{TAFRESHI18a,DUNLOP}, and therefore
it seems natural to take $\vec{f}_{\perp}$ to be the force that truly 
anchors a drop to a solid surface in the direction perpendicular to the 
surface. Second, the conceptual comparison of Eqs.~(\ref{difoat}) 
and~(\ref{difoasc1}) with the approaches of Refs.~\cite{TADMOR17,EXTRAND17} 
and with the Young-Dupr\'e work of adhesion must be done using $f_{\perp}$,
because in Refs.~\cite{TADMOR17,EXTRAND17} the force produced by pressure
was not taken into account.

To the best of our knowledge, Eq.~(\ref{difoat}) has been studied 
(with a different notation and for different situations) only in 
Refs.~\cite{TAFRESHI18a,TAFRESHI18b,TAFRESHI18c,TAFRESHI21,DUNLOP}. Our 
approach is essentially the same as that of 
Refs.~\cite{TAFRESHI18a,TAFRESHI18b,TAFRESHI18c,TAFRESHI21,DUNLOP}, the
only real difference being that for us the normal force of adhesion does 
not take the (Laplace) pressure into account, whereas in 
Refs.~\cite{TAFRESHI18a,TAFRESHI18b,TAFRESHI18c,TAFRESHI21} the normal force 
of adhesion is implicitly taken to be the same as the detachment force, 
which does include the contribution of the force produced by pressure.

Historically, the role played by the normal component of surface tension, 
$\gamma \sin \theta$, has been either neglected or just briefly 
mentioned~\cite{DEGENNES}. Some authors~\cite{PELLICER} have even 
dismissed it. For soft materials, however, it has been 
realized~\cite{LESTER,SHA1,SHA2,SHA3,SHA4,SHA5,SHA6,SHA7,SHA8} that 
$\gamma \sin \theta$ deforms the solid and produces a ``wetting ridge'' whose
height is of the order of $\gamma /G$, where $G$ is the 
shear modulus of the solid. Based on the work of Shanahan and 
co-workers~\cite{SHA1,SHA2,SHA3,SHA4,SHA5,SHA6,SHA7,SHA8}, 
Tadmor~\cite{TADMOR13} proposed a new formula
for the lateral retention force $\fp$. The main difference between 
our approach and that of Tadmor is that for us the {\it normal} component
of $\gamma$, $\gamma \sin \theta$, affects the
{\it normal} retention force, whereas for Tadmor it affects the 
{\it lateral} retention force. The point of view of the present paper is 
the same as that of Ref.~\cite{DEGENNES}: The normal component of surface 
tension is balanced out by a reaction force exerted by the solid
at the triple line. If the solid is hard, there is no significant wetting 
ridge. When it is soft (e.g., rubber or a coat of paint), 
$\gamma \sin \theta$ distorts the solid at the triple line and the 
results of Refs.~\cite{LESTER,SHA1,SHA2,SHA3,SHA4,SHA5,SHA6,SHA7,SHA8}
need to be used.

\subsection{The lateral advancing and receding works of adhesion}
\label{sec:arwa}

\begin{figure}[h!]
\begin{center}
              \epsfxsize=8cm
              \epsffile{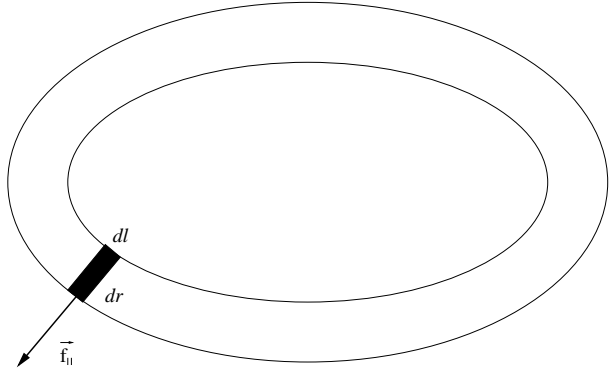}
\end{center}                
\caption{Contraction of the triple line by an amount $dr$. The
shape of the triple line is arbitrary.}
\label{fig:infiwork}
\end{figure}

Let us assume for simplicity that the only external force acting on the 
drop of Fig.~\ref{fig:stcca} is a force normal to the surface that is pulling
the drop away from the surface. As the
drop is deformed, the contact angle changes from its initial equilibrium
angle. When the contact angle becomes the receding angle $\theta _{\rm r}$, 
the triple line starts receding. Since $\fp$ is the force
that needs to be overcome to move the triple line, the work necessary
to move a segment of triple line $dl$ a distance $dr$ 
(see Fig.~\ref{fig:infiwork}) is given by~\cite{NOTE4} 
\begin{equation}
       d W_{\rm r} = \fpm \, dr = 
          \gamma (\cos \theta_{\rm r}-\cos \theta _{\rm Y}) dl\, dr \, .
\end{equation}

\noindent Hence, the work per unit area necessary to destroy an 
amount of contact area $dA=dr \, dl$ is
\begin{equation}
      {\sf w}_{\rm r}=\frac{d W_{\rm r}}{dA}=
       \gamma (\cos \theta_{\rm r}-\cos \theta _{\rm Y}) \, .  
          \label{wpeatocro}
\end{equation} 
To calculate the total work necessary to destroy 
an area ${\cal A}$ as the triple line recedes, we need to integrate
Eq.~(\ref{wpeatocro}) over the whole triple line,
\begin{equation}
     W_{\rm r}=\gamma \int_{\cal A} (\cos \theta _{\rm r} -\cos \theta _{\rm Y}) dA
           \, .
     \label{totalwork}
\end{equation}

If the external force acting on the drop pushes the drop towards the 
surface and the triple line expands, analogous equations hold by replacing
the receding angle with the advancing one,
\begin{eqnarray}
     {\sf w}_{\rm a}=\frac{dW_{\rm a}}{dA}=
       \gamma (\cos \theta_{\rm Y}-\cos \theta _{\rm a}) \, ,  
          \label{wpeatocroa}  \\
   W_{\rm a}=\gamma \int_{\cal A} (\cos \theta _{\rm Y} -\cos \theta _{\rm a}) dA
             \, .
     \label{totalworka}
\end{eqnarray}

Several comments are in order. First, similar to the Young-Dupr\'e work
of adhesion, Eqs.~(\ref{wpeatocro}) and~(\ref{wpeatocroa}) show that the
energy per unit area needed to create or destroy a patch of solid-liquid 
interface is a constant that depends only on surface tension and 
contact angles. Second, 
the work ${\sf w}_{\rm a}$ needed to create a solid-liquid patch by expanding 
the triple line is the same as the work ${\sf w}_{\rm r}$ needed to destroy
the same patch by contracting the triple line only when
the following relationship holds:
\begin{equation}
      \cos \theta _{\rm Y}= \frac{ \cos \theta _{\rm a} + \cos \theta _{\rm r}}{2}
\, .
   \label{eqneryea}
\end{equation}
Third, similar to the 
Young-Dupr\'e work of adhesion~\cite{SCHRADER}, the advancing and receding
works of adhesion do not take into account the energy needed to change
the shape of the drop. Only the energy needed to create or destroy an area of 
solid-liquid interface is taken into account. Fourth, 
conceptually, the main difference between the work of adhesion and
the lateral works of adhesion is that ${\sf w}_{\rm adhesion}$ is a
thermodynamic quantity that measures the energy needed
to detach a drop from a surface, whereas  ${\sf w}_{\rm a}$ and 
${\sf w}_{\rm r}$ quantify the energy dissipated by $\fp$ as the 
triple line moves on a surface. Fifth, in principle, we can 
use Eqs.~(\ref{totalwork}) and~(\ref{totalworka})
to calculate the total lateral work of adhesion needed to create or
destroy a solid-liquid contact area $A$. In practice, however,
we need to know how the contact angle changes along the triple line as
the contact area expands or contracts. For some special cases, we can
calculate such total lateral work exactly. For example, when the contact angle 
$\theta _{\rm r}$ ($\theta _{\rm a}$) remains constant as the
solid-liquid contact area contracts (expands), the total work
to destroy (create) an area $A$ is
\begin{eqnarray}
     W_{\rm r}=\gamma (\cos \theta _{\rm r} -\cos \theta _{\rm Y}) A \, ,
     \label{totalworkcs} \\
    W_{\rm a}=\gamma (\cos \theta _{\rm Y} -\cos \theta _{\rm a}) A \, .
     \label{totalworkacs}
\end{eqnarray}
It may be surprising that Eqs.~(\ref{totalworkcs}) and (\ref{totalworkacs}) 
mix up thermodynamic, equilibrium quantities (such as $\theta _{\rm Y}$) 
with metastable, non-equilibrium ones (such as $\theta _{\rm a}$ and 
$\theta _{\rm r}$). The apparent inconsistency is resolved by 
realizing that the way we calculated $W_{\rm a}$ and $W_{\rm r}$ uses
a purely mechanical approach (work done by the lateral
retention force), without invoking
equilibrium thermodynamics. In addition, one can always use the Young equation
to express $\cos \theta _{\rm Y}$ in terms of surface tensions and therefore
make $W_{\rm a}$ and $W_{\rm r}$ depend explicitly on non-equilibrium quantities.

\subsection{The lateral work of sliding}
\label{ref:lwoa}

When a drop slides on a surface at a low, 
constant velocity, the solid-liquid contact area has a 
quasi-rectangular shape~\cite{FURMIDGE,DUSSAN}, and as an amount of 
solid-liquid contact area is 
destroyed, the same amount is created. Hence, the work per unit
area necessary to slide the drop is
\begin{equation}
      {\sf w}_{\rm sliding} = {\sf w}_{\rm r}+{\rm w}_{\rm a} = 
      \gamma (\cos \theta_{\rm r}-\cos \theta _{\rm a})  \, .
         \label{wosliding}
\end{equation}
Equation~(\ref{wosliding}) was first introduced by 
Furmidge~\cite{FURMIDGE,NOTE3}. However, Furmidge assumed an expression for
the advancing and receding works of adhesion based on the Young-Dupr\'e
work of adhesion,
\begin{eqnarray}
      & {\sf w}_{\rm a}= -\gamma (1+\cos \theta_{\rm a}) \, ,   
       & \qquad [{\rm Ref}.~1] \, ,
             \label{awaF}\\
      & {\sf w}_{\rm r}= \gamma (1+\cos \theta_{\rm r})  \, , 
         &    \qquad [{\rm Ref}.~1] \, .
            \label{rwaF}
\end{eqnarray}
Although the sum of Eqs.~(\ref{awaF}) and~(\ref{rwaF}) does yield
the same result as Eq.~(\ref{wosliding}), it is problematic to assume 
that the Young-Dupr\'e equation can be applied freely to non-equilibrium
situations.

The quantities ${\sf w}_{\rm a}$, ${\sf w}_{\rm r}$ and 
${\sf w}_{\rm sliding}$ do not take
into account the energy lost due to viscous dissipation. However, in
situations where the capillary number is low, it would not be surprising
that viscous dissipation plays a minor role compared to the energy
dissipated by the lateral retention force.

\section{Experimental section}
\label{sec:exsect}

\begin{figure}[h!]
\begin{center}
              \epsfxsize=9cm
              \epsffile{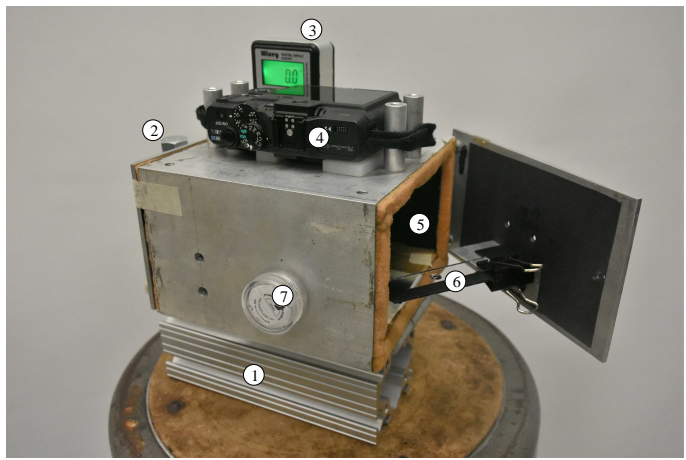}
\end{center}                
\caption{Experimental apparatus: {\bf 1}-platform; {\bf 2}-screw; 
{\bf 3}-inclinometer; {\bf 4}-top camera; 
{\bf 5}-side camera; {\bf 6}-PMMA sheet with drop; {\bf 7}-remote-controlled 
LED light. 
Inside the box (not shown in the picture) there is an LED
panel that provides the necessary illumination for the top camera.}
\label{fig:apparatus}
\end{figure}

Our experimental apparatus consists of a metallic 
box~\cite{LANGMUIR-RAFA} that sits on top of a platform, see 
Fig.~\ref{fig:apparatus}. 

\noindent Using a screw, the box can be inclined at any desired 
angle. Each turn of the screw changes the 
inclination of the box by around 0.6$^\circ$. An inclinometer 
is attached to the box to measure the tilt angle $\alpha$. The box
has two cameras placed on top and on the side, which
provide top and side views of the drops. On the door of the metallic box, 
a poly methyl methacrylate (PMMA) sheet 
(Optix, by Plaskolite) can be mounted such that, 
when a (distilled) water drop is placed 
on the sheet and the door is closed, the cameras have side and top views of the 
drop. Illumination for
the side camera is provided by a remote-controlled LED. Lighting for the
top camera is provided by an LED panel and an optical 
gradient~\cite{PODGORSKI}. The remote-controlled LED is used to set a 
common starting time in the videos of the cameras. To remove any remnants 
of their protective 
films, the PMMA sheets were initially washed with hot water and soap. 
Afterward, before each run, the PMMA sheets were cleaned
with 70\% isopropyl alcohol and paper tissue, 
and dried with lamplight~\cite{LANGMUIR-RAFA,AJP15}. To place the (distilled) 
water droplets on the PMMA sheet, we used a syringe from Hamilton. The 
volume of the water drops was 60~$\mu$L.

\begin{figure}[h!]
\begin{center}
              \epsfxsize=12cm
              \epsffile{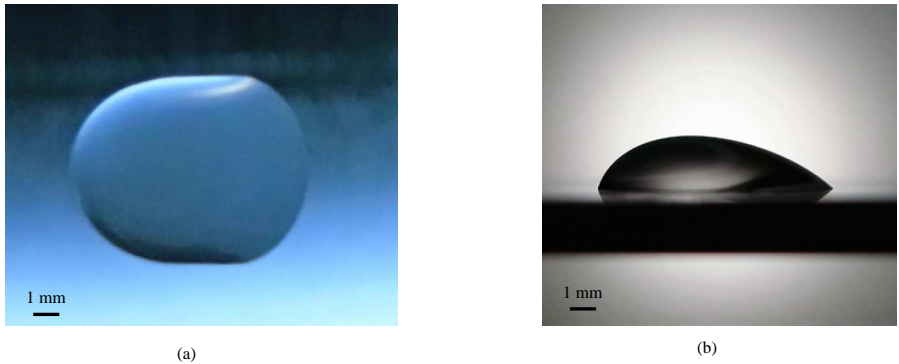}
\end{center}                
\caption{Top (a) and side (b) views of a drop moving with nearly 
constant velocity.}
\label{fig:recplan}
\end{figure}

For each run, we placed a drop on the PMMA sheet, closed the door, and
used the screw to tilt the box. After the drop started to slide down the 
surface, the inclination was slightly lowered to make the drops 
move with a slow, constant velocity. When this was achieved, the contact 
area of the drop had a quasi-rectangular shape, see Fig.~\ref{fig:recplan}.

We used the videos of the side camera and custom-made software to 
determine the position and speed of the receding edge of the drop. When
the speed was about constant and the contact area was quasi-rectangular, 
we used ImageJ~\cite{IMAGEJ} to obtain the contact angles 
$\theta _{\rm a}$ and $\theta _{\rm r}$ from 
the videos of the side-view camera, and the width
and the contact area from the videos of the top-view camera. Before the
box was tilted, we measured the equilibrium
angle $\theta_{\rm Y}$. In both theory and experiment, the angles are 
the apparent, macroscopic angles.

Our experimental apparatus is not able to detach drops, and therefore
we cannot directly measure $\vec{f}_{\perp}$. However, using the so-called 
Centrifugal Adhesion Balance (CAB), Tadmor~{\it et al.}~\cite{TADMOR17} 
were able to detach drops from surfaces. The CAB has
a centrifugal arm that rotates in a horizontal plane. At one end of the
arm, there is a chamber where a solid substrate can be placed. This chamber
can rotate around an axis that is orthogonal to the centrifugal 
rotation. As the centrifugal rotation proceeded and 
the centrifugal force increased, the chamber was tilted in such
a way that the external 
lateral force acting on the drop was always zero, whereas the external normal
force increased. Thus, essentially, the CAB provided an external
force that was ever increasing and normal to the surface. When the normal 
force was strong enough, the drop was 
detached from the surface, and the work of adhesion was calculated. We will
analyze the experiment of Ref.~\cite{TADMOR17} in the light 
of Eq.~(\ref{difoasc1}).

\section{Results and discussion}

\subsection{The normal retention force}
\label{sec:nfoad}


In Ref.~\cite{TADMOR17}, the value of $f_{\perp}$ was used
to obtain the Young-Dupr\'e work of adhesion as the
ratio of the pulling force to the length of the (circumferential) triple line,
\begin{equation}
    \frac{f_{\perp}}{2\pi r} = {\sf w}_{\rm adhesion} =\gamma (1+\cos \theta_{\rm Y})
                \, , 
         \hskip1cm [{\rm Ref}.~26].
         \label{perrwoaT1}
\end{equation}
In Ref.~\cite{EXTRAND17}, a different approach to the problem leads to
a different expression for the work of adhesion,
\begin{equation}
         \frac{f_{\perp}}{2\pi r} ={\sf w}_{\theta}=\pm \gamma \cos \theta_{\rm c}
                \, , 
         \hskip1cm [{\rm Ref}.~27],
         \label{perrwoaT1ext}
\end{equation}
where $\theta _{\rm c}$ is the critical angle at which the drop is 
detached from the surface, and
where the $+$ ($-$) sign corresponds to the case $\theta_{\rm c} <90^\circ$
($\theta_{\rm c}>90^\circ$). By contrast, when we apply 
Eq.~(\ref{difoasc1}) to the same situation, and assuming that we neglect
the force due to pressure, we obtain
\begin{equation}
      \frac{f_{\perp}}{2\pi r} = \gamma \sin \theta _{\rm c} \, ,
      \hskip1cm [{\rm this \ work}].
       \label{perrwhere}
\end{equation}
Clearly, Eq.~(\ref{perrwoaT1}), (\ref{perrwoaT1ext}) 
and~(\ref{perrwhere}) assign different values and meanings to one
and the same quantity, $f_{\perp}/2\pi r$.


The experiment of Ref.~\cite{TADMOR17} measured the values
of the detachment force $f_{\rm d}$ and radius $r$ of water
drops on
three different surfaces (hydrophobic silicon, microporous PTFE, and
hydrophilic glass). The values were compiled by Extrand in 
Ref.~\cite{EXTRAND17}, and are listed in 
Table~\ref{table:comparison}~\cite{NOTE5}. Using the experimental value of
$f_{\rm d}/2\pi r$, and assuming that the pressure force can be neglected
(i.e., assuming that $f_{\rm d}=f_{\perp}$),
we can use Eqs.~(\ref{perrwoaT1}), (\ref{perrwoaT1ext})
and (\ref{perrwhere}) to obtain the corresponding contact angles,
see Table~\ref{table:comparison}. In Table~\ref{table:comparison}, we 
also list the measured, apparent, macroscopic contact angles 
$\theta_{\rm measured}$.

\begin{center}
\begin{table}[!h]
\begin{center}
\caption{Comparison of the predictions of 
Eqs.~(\ref{perrwoaT1}), (\ref{perrwoaT1ext})
and (\ref{perrwhere}).} 
\begin{tabular}{c| c c c c c c c c c c c} 
\hline\hline
\multirow{2}{*}{Solid surface}  & $\frac{f_{\rm d}}{2\pi r}$
& $\theta_{\rm Y}$$^{(\circ)}$ 
& $\theta_{\rm c}$$^{(\circ)}$
& $\theta_{\rm c}$$^{(\circ)}$ & $\theta_{\rm measured}^{\ (\circ)}$  \\ 
& (mJ/m$^2$)
& ({\rm Ref}.~[26])   
& ({\rm Ref}.~[27])  
& (this work)  & ({\rm Ref}.~[26])   \\
\hline
\hline 
Hydrophobic silicon & $50$ & $107$ & $46$ & $44$ & $81$  \\ [1ex]
Microporous PTFE & $26$ & $130$ & $111$ & $159$ & \ \ $123$ $^*$  \\ [1ex]
Hydrophilic glass & $72$ & $90$ & $0$ & N/A  & $20$  \\ 
\hline\hline 
\multicolumn{6}{l}{$^*$ \small{This value is a rough estimate based on
Fig.~9 of {\rm Ref}.~[26].}}
\end{tabular}
\label{table:comparison}
\end{center}
\end{table}
\end{center}

For hydrophobic silicon, our critical contact angle ($44^\circ$) is very close 
to that of Extrand ($46^\circ$). The value of
$\theta _{\rm Y}$ that comes out of Eq.~(\ref{perrwoaT1}) is, however,
much larger $(107^\circ)$. All of those angles are far from the apparent, 
measured
angle ($81^\circ$). For mPTFE, all the contact angles are again far from 
the (rough estimate of the) apparent contact angle, and ours is the
farthest of all. The most dramatic differences occur 
for hydrophilic glass. When a
drop of water was detached from a glass surface, a nanometric layer of water
was left behind, and Tadmor~{\it et al.}~concluded that
they were ``in practice separating water from water''~\cite{TADMOR17}. As
is well known~\cite{DAVIES}, when 
a liquid is separated from itself,
the relevant work is the work of cohesion, which according to 
Eq.~(\ref{woalfl}) is ${\sf w}_{\rm cohesion}=2 \gamma$. However, the 
experimental value of Ref.~\cite{TADMOR17} is
71.3~${\rm mJ}/{\rm m}^2\approx \gamma$. To match this value of
${\sf w}_{\rm adhesion}$, a Young contact angle of $90^{\circ}$ is needed,
which seems too high for water on glass~\cite{EXTRAND17}, even
when the increase of the contact angles due to the large value of the 
effective acceleration of gravity is 
taken into account. Extrand's
approach, Eq.~(\ref{perrwoaT1ext}), yields a critical angle 
of $0^{\circ}$. Our approach uses the work of cohesion, and therefore does 
not yield any contact angle.

It is clear from Table~\ref{table:comparison} that the predictions of 
all three approaches do not match the apparent, measured 
angles. Tadmor~{\it et al.}~\cite{TADMOR17} have argued that the mismatch 
occurs because the true nanoscopic Young contact angle is different from
the macroscopic, apparent contact angle. In our approach, however, it is 
possible to account for the apparent measured angles by including the 
effect of the force due to pressure. As mentioned in Sec.~\ref{sost}, 
Eq.~(\ref{perrwhere}) does not include the effect of the normal reaction 
force $f_{\rm p}$ on the liquid-solid contact area that is due to the 
pressure of the drop. In general, calculating the force due to pressure
is difficult, because it involves the Laplace pressure. However, for drops 
that are detached without a neck formation (like mPTFE),
Farhan and Tafreshi~\cite{TAFRESHI18a} were able to estimate the pressure 
force to be $f_{\rm p}= \gamma \pi r$, where $r$ is the radius
of the (circular) contact area. Because the pressure force points away from
the surface, it combines with the detachment force $f_{\rm d}$ to produce 
the normal force of adhesion, $f_{\perp}=f_{\rm d}+f_{\rm p}$. Hence,
\begin{equation}
\gamma \sin \theta _{\rm c}= \frac{f_{\perp}}{2\pi r} = \frac{f_{\rm d}}{2\pi r} +
\frac{\gamma}{2} \, ,
\end{equation}
which yields~\cite{TAFRESHI18a} a value of $\theta _{\rm c}=121^\circ$, very
close to the apparent contact angle ($123^\circ$) of
Table~\ref{table:comparison}. If we
took the effect of $f_{\rm p}$ into account, for mPTFE 
Tadmor's approach would yield a Young angle of $98^\circ$, and 
Extrand's approach would yield a contact angle of $149^\circ$.

As can be seen in the movie of the Supporting Information of
Ref.~\cite{TADMOR17}, the rotation of the CAB produces wild vibrations
on the drops. Hence, the values of the contact angles and the detachment
force are subject to a systematic error, and the comparison
of the contact angles of Table~\ref{table:comparison} cannot lead to
firm conclusions as to which approach best describes the detachment of a
drop. Without systematic-proof data, it is more useful to 
compare the conceptual foundations of the three approaches.

Conceptually, the main differences between Tadmor's approach and ours 
are the following: First, in Tadmor's approach, thermodynamic 
equilibrium quantities such as ${\sf w}_{\rm adhesion}$ and $\theta _{\rm Y}$ are 
measured in a seemingly non-equilibrium situation, whereas we use a 
completely mechanical approach without any recourse to 
equilibrium thermodynamics. Second, the data of Ref.~\cite{TADMOR17} 
has a dependence on the receding, critical angle~\cite{NOTE2}, and 
therefore needs to be described by
a formula that involves such angle instead of $\theta_{\rm Y}$. Third,
when water is separated from water, we use the 
work of cohesion rather than the Young-Dupr\'e work of adhesion. Fourth,
Eq.~(\ref{perrwoaT1}) is obtained by a loose analogy with Tate's law, whereas 
we have provided a thorough derivation of Eq.~(\ref{perrwhere}).

Conceptually, the main differences between Extrand's approach and ours are 
the following: First, Extrand assumes that $f_{\perp}$
is equal to a force $f_{\rm s}$ that is parallel to the surface and
that is given in terms of the component of surface tension that is
parallel to the surface, whereas in our approach
$f_{\perp}$ is obtained in terms of the component of surface tension that
is perpendicular to the surface. Second, as long as we assume that
the contact angle $\theta _{\rm c}$ remains constant along the circumferential
triple line (as Extrand seems to assume), the force $f_{\rm s}$ parallel to the
surface that would arise from integrating Eq.~(\ref{larfodlt}) along the
triple line would be zero, whereas Extrand estimated that
$f_{\perp}=f_{\rm s}= 2\pi r \gamma \cos \theta _{\rm c}$. 

Conceptually, our approach is very much the same as that 
of Farhan and Tafreshi~\cite{TAFRESHI18a}, the main difference being 
simply a matter of definition:
For us the normal retention force is not the same as the actual detachment
force, our normal retention force is the sum of the detachment force
and the force due to pressure.

Because the experimental values of Table~\ref{table:comparison} suffer
from a known systematic (vibration), the only way to elucidate 
the proper way to describe the normal retention force is by 
performing new experiments that avoid the vibrations of the CAB. Such 
experiments can be easily performed using the same technique of 
Refs.~\cite{TAFRESHI18a,TAFRESHI18b,TAFRESHI18c,TAFRESHI18d}, but
substituting the fibrous surface by a flat solid surface. A ferrofluid
droplet can be detached from a flat surface, and all the relevant experimental
quantities can be easily measured.

\subsection{The advancing and receding works of adhesion}
\label{sec:lwoad}

Table~\ref{table:exdata} lists our experimental data for drops sliding
down a PMMA surface. 
\begin{center}
\begin{table}[!h]
\begin{center}
\caption{Experimental data. All errors are statistical (standard
deviation of the mean).}
\begin{tabular}{c c c c| c c c c c c c c} 
\hline\hline
\multicolumn{4}{c}{Angles}  & 
              \multicolumn{1}{c}{Speed} & 
              \multicolumn{1}{c}{Area} & 
              \multicolumn{1}{c}{Width} &    \\
\hline

$\alpha \ ^{(\circ)}$  & $\theta_{\rm a} \, ^{(\circ)}$ & 
$\theta _{\rm r} \, ^{(\circ)}$ & $\theta _{\rm Y} \, ^{(\circ)}$  & 
$v \, ({\rm mm/s})$ &
$A \, ({\rm mm}^2)$ & $w \, ({\rm mm})$ \\
\hline
$24.1\pm 0.2$ & $72.8 \pm 0.5$ & $39.4 \pm 0.4$ & $60.7 \pm 0.4$ & 
$0.79 \pm 0.07$ & 
$54.8 \pm 0.2$ & $7.1\pm 0.2$  \\

\hline\hline 
\end{tabular}
\label{table:exdata}
\end{center}
\end{table}
\end{center}

\noindent The Young, 
equilibrium angle was measured when the PMMA surface was placed horizontally, 
whereas the rest of the quantities were measured when the surface was tilted
at an angle $\alpha$ and the drop was moving at
a nearly constant speed with a quasi-rectangular triple line. In our
experiments, we used 21 different PMMA sheets, and did 3 runs with
each sheet, for a total of 63 runs.

By plugging the data of Table~\ref{table:exdata} into 
Eqs.~(\ref{wofa}), (\ref{wpeatocro}), (\ref{wpeatocroa}) 
and~(\ref{wosliding}), we obtain the results of 
Table~\ref{table:exresults}. 


\begin{center}
\begin{table}[!h]
\begin{center}
\caption{Experimental results.}
\begin{tabular}{c c c c c c c c c c c c} 
\hline\hline 
              \multicolumn{4}{c}{Works of adhesion (mJ/m$^2$)}   \\
\hline
${\sf w}_{\rm a}$ & ${\sf w}_{\rm r}$ &
${\sf w}_{\rm sliding}$ & ${\sf w}_{\rm adhesion}$  \\
\hline
$13.9 \pm 0.7$ &  $20.4\pm 0.5$ & $34.3 \pm 0.9$ & $107.2\pm 0.4$  \\

\hline\hline 
\end{tabular}
\label{table:exresults}
\end{center}
\end{table}
\end{center}


\noindent We can draw several conclusions from the results 
of Table~\ref{table:exresults}. First, the work of sliding is somewhat 
larger than the advancing and receding works of adhesion. Second, the 
Young-Dupr\'e work of adhesion is significantly larger than any of the 
lateral works of adhesion. Third, the result that  
${\sf w}_{\rm a}<{\sf w}_{\rm sliding}<{\sf w}_{\rm adhesion}$ is 
consistent with the experimental fact that the force necessary to 
spread a drop on a surface is smaller than the
force necessary to slide the drop on the surface, which in turn is 
much smaller than the force needed to detach the drop from the 
surface~\cite{TADMOR17,GRIEGOS1,GRIEGOS2,GRIEGOS3}. Fourth, the advancing 
and receding works of adhesion are different. Although it is possible 
that ${\sf w}_{\rm a}$ and ${\sf w}_{\rm r}$ are truly different, 
it is also possible that their actual values are the same. The reason
is the following: It is 
believed that the apparent, equilibrium angles of sessile drops often lie in 
between the true Young contact angle and the advancing contact angle, due
to the way the drop is deposited on the surface. Because overestimating 
$\theta _{\rm Y}$ leads to overestimating ${\sf w}_{\rm r}$ and 
underestimating ${\sf w}_{\rm a}$, the difference between ${\sf w}_{\rm a}$ and 
${\sf w}_{\rm r}$ may be simply due to not having measured the 
true Young contact angle. If ${\sf w}_{\rm a}$ and ${\sf w}_{\rm r}$ were 
actually equal, Eq.~(\ref{eqneryea}) would provide the 
true Young equilibrium angle. For our system, the Young angle predicted by
Eq.~(\ref{eqneryea}) is $57.7^\circ$, which is smaller than, but very
close to, the apparent Young angle we measured ($60.7^\circ$). Hence, although
it is a speculative proposal, it may well be that the true Young contact angle
is given by Eq.~(\ref{eqneryea}), and that ${\sf w}_{\rm a}={\sf w}_{\rm r}$.

The capillary number, defined as ${\rm Ca}=\frac{\eta v}{\gamma}$, 
where $\eta$ is the viscosity of the liquid and $v$ its speed, measures
the importance of viscous forces relative to capillary forces. In our 
experiment, ${\rm Ca}\sim 10^{-5}$, which suggests that in our experiment
the dissipation by the lateral retention force (quantified by the
sliding work of adhesion) dominates 
over viscous dissipation. 

To calculate how much gravitational potential
energy is dissipated by $\fp$, we are going to compare the gravitational
potential energy of the drop with the energy dissipated by $\fp$, or,
equivalently, we are going to compare the work per unit 
area done by the weight with the sliding work 
of adhesion. To calculate the work done by the weight, let us consider a 
drop of 
quasi-rectangular contact area $A$, length $L$, and width $w$ that is 
sliding down the incline with constant velocity. When the drop travels a 
distance $L$, its center
of mass drops a height $h=L\sin (\alpha)$, and therefore the work done
by the weight is
\begin{equation}
         W_{\rm g}=mgh = mg L \sin (\alpha) \, .
\end{equation}
Since $A=Lw$, the work per unit area done by the weight is
\begin{equation}
     \frac{W_{\rm g}}{A} =  \frac{mgL\sin(\alpha)}{A} = 
            \frac{mg\sin (\alpha)}{w} \, .
       \label{wpadbyw}
\end{equation}
By plugging the values of Table~\ref{table:exdata} into Eq.~(\ref{wpadbyw}),
we obtain the following experimental value: 
\begin{equation}
       \frac{W_{\rm g}}{A}= (34 \pm 1)~{\rm mJ/m}^2 \, , \label{enagree1} 
\end{equation}
which agrees within 1\% with the value of ${\sf w}_{\rm sliding}$ in
Table~\ref{table:exresults}. Hence,
viscous dissipation can be neglected in our experiment, since
all the gravitational potential energy is dissipated by the lateral
retention force. 

Theoretically, the condition that $\frac{W_{\rm g}}{A}$ equals 
${\sf w}_{\rm adhesion}$ is equivalent to the condition that the
lateral retention force equals the component of the weight parallel to the 
surface,
\begin{equation}
      \fpm = mg\sin (\alpha) \, .
        \label{fpewain}
\end{equation} 
Hence, Eq.~(\ref{fpewain}) is another way to check that viscous dissipation
can be neglected. By 
plugging the values of Table~\ref{table:exdata} into Eqs.~(\ref{fpewain})
and~(\ref{trf}), we obtain the following experimental agreement:
\begin{eqnarray}
       \fpm = \gamma w (\cos \theta _{\rm r}-\cos \theta _{\rm a})= 
               (244 \pm 8)~\mu{\rm N} \, ,  \label{forceagree1}\\
       mg\sin(\alpha) = (240 \pm 2)~\mu{\rm N} \, . \label{forceagree2}
\end{eqnarray}
Hence, both energy- and force-wise, viscous dissipation
plays no significant role in our water-PMMA system. 

Since the gravitational potential energy is dissipated by the
lateral retention force, viscoelastic 
dissipation~\cite{LESTER,SHA1,SHA2,SHA3,SHA4,SHA5,SHA6,SHA7,SHA8}
can also be neglected in our system. This is not surprising, since the height 
of the ``wetting ridge'' of the water-PMMA system is of the order of 
$\gamma /G \sim 42$~pm, certainly negligible on a macroscopic scale.

Although we have only studied water droplets sliding on a PMMA surface,
we would like to note that
Furmidge's experimental values~\cite{FURMIDGE} are consistent with
Eq.~(\ref{fpewain}), and therefore viscous dissipation can also be neglected 
for the solid-liquid pairs considered by Furmidge~\cite{FURMIDGE}.

From the discussion above, it should be clear that our claim that the 
gravitational potential energy is dissipated (mostly) by the lateral 
retention force 
rests on the assumption that Eq.~(\ref{trf}) provides the exact lateral
retention force. Such assumption holds only when the
advancing (receding) contact angle is constant along the advancing (receding)
edge of the drop. However, in Ref.~\cite{ANTONINI} it has been shown that 
neglecting the variation of the contact angle along the triple line may lead 
to an erroneous value of the lateral retention force. Hence, to check that
Eq.~(\ref{trf}) provides a very good approximation of the actual
lateral retention force, one needs a measurement of the
variation of the contact angle along the triple 
line. Although our experimental apparatus is not able to measure such
variation, the so-called IBAFA methodology of Ref.~\cite{ANTONINI} could 
measure it, thereby
characterizing how accurately Eq.~(\ref{trf}) describes
$\fpm$ and how much gravitational potential energy is dissipated by $\fpm$.

We would like to note that the typical size of a 60~$\mu$L drop (around 7~mm) 
is larger than the capillary length of water (around 2.7~mm), and therefore 
in our experiment gravity dominates
over surface tension. This makes 60~$\mu$L water drops sliding on
an incline have a quasi-rectangular shape, a situation where we expect
Furmidge's expression for the lateral retention force to be accurate. However,
small drops whose characteristic length is less than the capillary length
will have 
a more circular-like triple line, Furmidge's expression may not be
so accurate, and viscous dissipation may be more important for them than
for larger drops.

\section{Conclusions}
\label{sec:conclusions}

We have calculated the normal capillary retention force
$\vec{f}_{\perp}$ on a drop that is in contact with a solid surface. We have
seen that, essentially, $\vec{f}_{\perp}$ is determined by the component of 
surface tension that is perpendicular to the solid substrate. We have 
used the expression for $\vec{f}_{\perp}$ to provide a new theoretical
description of a recent experiment on the work of 
adhesion~\cite{TADMOR17,EXTRAND17}, and concluded that the 
normal retention force per 
unit length of triple line is different from the Young-Dupr\'e
work of adhesion. We have also compared the conceptual foundations
of $\vec{f}_{\perp}$ with the approaches of 
Refs.~\cite{TADMOR17,EXTRAND17}. However, because of the wild vibrations 
of the CAB~\cite{TADMOR17}, further experimental work is needed
to elucidate which approach is more appropriate to describe the 
detachment of a drop from a surface. We have proposed that separating
ferrofluid droplets from a flat surface using the experimental apparatus of 
Refs.~\cite{TAFRESHI18a,TAFRESHI18b,TAFRESHI18c,TAFRESHI18d}
can provide the necessary data to elucidate the proper way to describe
the normal force of adhesion.

By calculating the work done by the lateral retention force, we have 
obtained the energy needed to contract (expand) the triple line
on a surface, and have identified the resulting energy
with the advancing (receding) work of adhesion ${\sf w}_{\rm a}$ 
(${\sf w}_{\rm r}$). We have used the expressions for 
${\sf w}_{\rm a}$ and ${\sf w}_{\rm r}$ to re-derive~\cite{FURMIDGE} the
expression for the sliding work of adhesion, ${\sf w}_{\rm sliding}$.

We have seen that the main similarity between the Young-Dupr\'e and
the lateral works of adhesion is that all of them represent work per unit
area that is equal to a constant that depends only on surface tension and
contact angles. Another similarity is that those works do not quantify 
any change in the shape of the drop that may occur, or the energy needed
to create or destroy the concomitant liquid-vapor contact area. The 
main difference between them is that whereas the Young-Dupr\'e work of adhesion 
quantifies the energy needed to create/destroy the solid-liquid
contact area by attaching/detaching a drop to/from a surface,
the lateral works of adhesion quantify the energy necessary to create/destroy
the solid-liquid contact area by moving the triple line along the surface.
Conceptually, the main difference is that the Young-Dupr\'e work of adhesion
is a thermodynamic equilibrium quantity, whereas the lateral works of adhesion 
are based on apparent angles and describe dynamical, non-equilibrium situations 
in which the triple line is moving on the surface. 

We have argued that when the capillary number is low and the drop
is moving at constant speed, viscous dissipation can be neglected compared 
to the energy dissipated by the lateral retention force. Because neglecting 
viscous dissipation in our system is equivalent to assuming that Furmidge's 
expression for the lateral retention force is exactly true, to know how much
gravitational potential energy is truly dissipated by $\fpm$, we need to
know how truly Furmidge's expression describes $\fpm$. We have proposed
that an experiment similar to that of Ref.~\cite{ANTONINI} should be able
to determine how much gravitational potential energy is truly dissipated
by the lateral retention force.

Our results are valid for flat surfaces. When the surface is not flat, or
if we considered a fiber~\cite{TAFRESHI18a,TAFRESHI18b,TAFRESHI18c,TAFRESHI18d},
the general principles are the same, although the integrations needed to
calculate the forces are more 
complicated~\cite{TAFRESHI18a,TAFRESHI18b,TAFRESHI18c,TAFRESHI18d}.

Finally, we would like to stress that both $\fp$ and $\vec{f}_{\perp}$ have 
a capillary (surface tension) origin. Hence, when other non-capillary forces
(e.g, viscous and viscoelastic forces) cannot be neglected, our results
would need to be modified.

\section*{CRediT authorship contribution statement}

Rafael de la Madrid: Conceptualization, data acquisition, writing-review \&
editing. Huy Luong: Design and construction of experimental apparatus. Jacob
Zumwalt: Data analysis.

\section*{Acknowledgments}

Support from a Lamar CICE grant is gratefully acknowledged. One of the 
authors (RdlM) was supported by a Lamar COAS fellowship. Correspondence 
with Amir Fatollahi is gratefully acknowledged.




\begin{thebibliography}{99}
\bibitem{NOTE1} Throughout the paper, we will use upper case $W$ to 
denote total work,
and lower case ${\sf w}$ to denote work per unit of area.
\bibitem{NOTE4} To simplify the derivation, we are going to assume that
the contact angles are smaller than 90$^\circ$. When some of (or all) the contact
angles are greater than 90$^\circ$, we need to change the signs in the final
formulas appropriately to make the works of adhesion positive, 
because it takes energy to move the triple line.
\bibitem{NOTE3} It should be noted that Furmidge~\cite{FURMIDGE} did not 
explicitly use the nomenclature of advancing, receding and sliding
works of adhesion.
\bibitem{NOTE5} The values listed by Extrand~\cite{EXTRAND17} are 
slightly different from the experimental values listed in 
Ref.~\cite{TADMOR17}. However, to keep the discussion simple, we will ignore
such difference because it does not affect our conclusions.
\end{thebibliography}
\end{document}